# Electric-field-controllable high-spin SrRuO$_3$ driven by a solid ionic junction

Jingdi Lu[1#], Liang Si[2, 3#], Xiefei Yao[1#], Chengfeng Tian[1#], Jing Wang[4#], Qinghua Zhang[5#], Zhengxun Lai[6], Iftikhar Ahmed Malik[1], Xin Liu[1], Peiheng Jiang[2], Kejia Zhu[5], Youguo Shi[5], Zhenlin Luo[7], Lin Gu[5], Karsten Held[3], Wenbo Mi[6], Zhicheng Zhong[2*], Ce-Wen Nan[4] and Jinxing Zhang[1*]

1. Department of Physics, Beijing Normal University, 100875 Beijing, China
2. Key Laboratory of Magnetic Materials and Devices, Ningbo Institute of Materials Technology and Engineering, Chinese Academy of Sciences, Ningbo, 315201 China
3. Institut für Festkörperphysik, TU Wien, Wiedner Hauptstraße 8-10, 1040 Vienna, Austria
4. School of Materials Science and Engineering, Tsinghua University, 100084 Beijing, China
5. Institute of Physics, Chinese Academy of Science, 100190 Beijing, China
6. Tianjin Key Laboratory of Low Dimensional Materials Physics and Preparation Technology, School of Science, Tianjin University, Tianjin 300354, China
7. National Synchrotron Radiation Laboratory & CAS Key Laboratory of Materials for Energy Conversion, and Department of Physics, University of Science and Technology of China, Hefei 230026, China

#These authors contributed equally to this work.
*E-mail: zhong@nimte.ac.cn; jxzhang@bnu.edu.cn

**Controlling magnetism and spin structures in strongly correlated systems by using electric field is of fundamental importance but challenging. Here, a high-spin ruthenate phase is achieved via a solid ionic chemical junction at SrRuO$_3$/SrTiO$_3$ interface with distinct formation energies and diffusion barriers of oxygen vacancies, analogue to electronic band alignment in semiconductor heterojunction. Oxygen vacancies trapped within this interfacial SrRuO$_3$ reconstruct Ru-4$d$ electronic structure and orbital occupancy, leading to an enhanced magnetic moment. Furthermore, an interfacial magnetic phase can be switched reversibly by electric-field-rectifying oxygen migration in a solid-state ionic gating device, providing a framework for atomic design of functionalities in strongly correlated oxides using a way of solid chemistry.**

Controlling magnetism and low-dimensional spin textures in strongly correlated systems by electric field is fundamentally significance due to their potential applications in information processing with low-power consumption [1-4]. Over past decades, spin degree of freedom have been effectively controlled by electric field in magnetoelectrics and multiferroic heterostructures [5-11]. However, a combination and control of these ferroic orderings with advanced characterizations of those coupled behaviors is still fundamentally challenging [1,3]. Therefore, it is desirable to design new correlated electron systems with an atomic precision by engineering of material chemistries and architectures [3]. Transition metal oxides (TMOs) with strong *d*-electron correlations and spin-orbit coupling (SOC) may provide an effective platform towards emergent magnetism with controllable spin textures [12-20].

SrRuO$_3$ (SRO) is a unique 4*d* TMO with a coexistence of SOC and itinerant ferromagnetism [21], leading to a large variety of physical behaviors which have been experimentally observed such as high-spin states driven by crystal orientations [22-24], magnetic skyrmions due to the broken symmetry [25-28], and metal-insulator transition [29-31], etc. More importantly, emergent physical behaviors have been theoretically proposed in SRO recently [13,15,32,33], triggering a broad attention to discover the potential quantum states in this material. Among these theories, heavy electron doping in SRO seems to be effective to further enhance magnetism and change electronic structure [15], which is, however, experimentally challenging in this oxide with a metallic ground state. Although a high concentration of the oxygen vacancy ($V_o$) is equivalent to electron doping, it is hard to be achieved in bulk SRO. This gives a strong impetus to explore an alternative pathway to build upon the above physical scenario. Inspired by electrons rectifying at solid-state junctions, such as the well-known diode at *p-n* junctions [34], a solid ionic chemical junction may be used to rectify $V_o$ by using its chemical discontinuity across the interface [35].

Atomically flat SRO thin films with accurate thicknesses of 5 uc, 10 uc, 30 uc and 50 uc (where uc is unit-cell) were grown on (001)-oriented SrTiO$_3$ (STO) substrates with TiO$_2$-termination by pulsed laser deposition (PLD) with *in-situ* reflection high-energy electron diffraction (RHEED). Through a synchrotron-based X-ray diffraction (SXRD) study, a set of (002)-oriented peaks in Fig. 1 (a) reveal SRO thin films with a c-axis lattice constant of 3.967 Å (thickness up to 50 uc). The epitaxial growth of SRO thin films is shown in Fig. S1. High-resolution scanning transmission electron microscopy (STEM) in Fig. 1 (b) indicates a high-quality and dislocation-free epitaxy for the SRO thin films. The out-of-plane and in-plane lattice structures in STEM (Fig. 1 (c)) illustrates that there is an in-plane compressive strain (~ -0.45 %) without any relaxation and phase separation across the whole SRO thin films, which are consistent with the results obtained from the x-ray reciprocal space mapping (RSM) (Fig. 1 (d)). The monoclinic structure of the SRO thin films [36] have been further revealed by the x-ray RSM. The details of growth and measurements can be seen in supplementary methods.

The ferromagnetic Curie temperature ($T_C$) was measured as a function of thickness by temperature-dependent magnetization and resistivity ($M$-$T$ and $R$-$T$) [21,37]. As shown in the Fig. 2 (a) and (b), $T_C$ ~153 K is observed for the 5 uc SRO thin film, while another $T_C$ ~125 K appears with the thickness increasing from 10 uc to 50 uc, the decreased $T_C$ is ascribed to the fully strained monoclinic structure as studied previously [36,38]. The thickness-dependent magnetic hysteresis loops along the pseudo [001] axis were measured by superconducting quantum interference device (SQUID) (Fig. 2 (c)). Upon increasing the thickness from 5 uc to 10 uc, a step-like hysteresis with two coercive fields appears, which are more obvious in the 30 uc and 50 uc thin films. For the 5 uc SRO, the out-of-plane saturated moment ($M_s$) is ~2.18 $\mu_B$/formula unit (f. u.), which is higher than the one of low-spin state of SRO (~1.1 – 2.0 $\mu_B$/f. u. in experiments and theories, 2.0 $\mu_B$/f. u. from full Hund's rule) [39-42]. As the thickness increases, the $M_s$ decreases to ~2.06, ~1.75 and ~1.28 $\mu_B$/f. u. for 10, 30, and 50 uc SRO, respectively, indicating that the SRO at the interface contributes to the enhancement of $M_s$ as shown in insets of Fig. 2 (c) and (d). The magnetic domain switching, corresponding to the hysteresis, has been characterized by using low-temperature magnetic force microscopy (MFM) at 10 K. For the 5 uc SRO, a mono-domain switching occurs (Fig. S2). However, for the SRO thin film with a thickness above 10 uc, there is a distinctive switching with two coercive fields at low/high magnetic field (Fig. S3), which further indicates that the step-like magnetic hysteresis could be attributed to a separation of magnetic phases: interfacial-SRO (high $T_C$/moment) and top-SRO (low $T_C$/moment). The hysteresis behavior as a function of top-SRO thickness is simulated in Fig. S4, which is consistent to our experimental observations in Fig. 2 (c). The details of measurements for magnetic/electrical characterizations are given in supplementary methods and Fig. S5.

In these fully-strained SRO thin films without lattice relaxation, slight alternation of chemical structure at the interface may play a key role on the enhanced magnetization [43]. As seen from bright-field STEM in Fig. 3 (a) and (b), the interfacial-SRO phase presents more $V_o$ which is negligible in the top-SRO. Depth-profiling X-ray photoelectron spectroscopy (XPS) with *in-situ* etching of surface layers without damaging underlying layers (see Fig. 3 (c) for the schematics) shows the valence states of Ru at various etching depths for the 50 uc SRO in Fig. 3 (d). The binding energy of Ru $3d_{3/2}$ remains at ~282.82 eV while etching away up to 30 uc. However, it starts to dramatically decrease by ~1.3 eV when the etching depth reaches about 45 uc, indicating that $Ru^{3+}$ appears at the interfacial-SRO [44]. At all conditions, there are no change for the binding energies of Sr $3d$ from the top-SRO to interfacial-SRO and then STO (see Fig. S6). Therefore, the appearance of $Ru^{3+}$ at the interfacial-SRO is induced by the $V_o$ based on the above analysis of STEM and XPS results [44]. The details of measurements for chemical characterizations are given in supplementary methods.

The microscopic origin of this $V_o$-dominant interfacial-SRO phase with enhanced $M_s$ can be understood by calculating formation energy ($E_{OF}$) and diffusion barrier ($E_b$) of the $V_o$ in both SRO and STO. The $E_{OF}$ is defined as

$$E_{OF} = E_{SC}(V_o) - E_{SC} + E(O_2)/2,$$

for both SRO and STO in density functional theory (DFT). Here, $E_{SC}(V_o)$ and $E_{SC}$ are the energies of supercells with and without a $V_o$, respectively, and $E(O_2)$ is the energy of a single $O_2$ molecule. The $E_{OF}$ of bulk SRO is ~4.7 eV, which is ~2 eV lower than that of bulk STO, indicating that the $V_o$ prefers to be on the SRO side. Near the SRO/STO interface, the $E_{OF}$ of a single $V_o$ in SRO is ~4 eV, lower than that in STO (Fig. 3 (e)). Therefore, we expect that $V_o$ migrates into the interfacial-SRO phase when SRO is in contact with STO. In the Fig. S7, we further calculated the diffusion energy barrier $E_b$ of $V_o$. In SRO, $E_b$ is ~1.4 eV, which is much higher than the one in STO (~0.6 eV). While for STO, such a barrier height permits $V_o$ diffusion and in SRO it prevents $V_o$ diffusing further into the SRO film [45]. As a result, the $V_o$ is trapped at the interfacial-SRO phase as visualized in Fig. 3 (e), and forms a $V_o$ diode, in a similar way as electron diffusion in a conventional electrical diode.

For understanding the role of $V_o$ in the enhanced $M_s$ at the interfacial-SRO phase, we construct a SRO (5 uc)/STO (4 uc) heterostructure, where one oxygen atom is removed between the 2nd and 3rd RuO$_2$ layers to model the $V_o$ (see Fig. 4 (a) and (b)). There is a substantial spin and orbital reconstruction of the Ru $d$-electrons near the $V_o$. The occupation of Ru $d_{z^2}$-orbit arises from the fact that i) $V_o$ provides abundant electrons, and ii) it breaks the neighboring RuO$_6$ octahedron and changes the crystal-field splitting. The latter shifts the energy of $d_{z^2}$-orbit down by forming bonding and anti-bonding states. This orbital reconstruction is pivotal for turning the $4d^4$ low-spin state ($S = 1$, ~1.5 $\mu_B$/Ru, far away from $V_o$) into $4d^5$ high-spin state ($S = 3/2$, ~2.3 $\mu_B$/Ru, neighboring $V_o$) as shown in Fig. 4 (c) and (d), being compatible with our experimental results (Fig. 2 (d)). Thus, the interfacial-SRO phase with high spin can be defined as $V_o$-SRO phase. The computational details are attached in the supplementary materials.

The interfacial high-spin SRO phase originates predominately from the $V_o$ contributions, which could be naturally controllable in this $V_o$ diode across the solid $V_o$-SRO/STO chemical junction by an electric field [46]. The temperature-dependent transport measurements with *in-situ* magnetic-domain characterizations were performed in the 10 uc SRO thin film under the application of electric field as schematically shown in Fig. 5 (a). The details on measurements can be seen in supplementary methods. A reversible control of the $V_o$-SRO ($T_C$ is ~153 K) was achieved by electric-field-driven uni-directional $V_o$ migration from SRO to STO as shown in Fig. 5 (b), demonstrating that the $V_o$-SRO with high spin can be erased and rebuilt by applying electric field. With a magnetic bias between the coercive fields of $V_o$-SRO and top-SRO, the soft $V_o$-SRO will switch firstly and serve as an effective magnetic field ($H_{eff}$). Magnetic domain switching with and without the application of electric field has been performed under a magnetic bias of +0.75 T, which is lower than the coercive field of top-SRO in the 10 uc film (~1 T, see Fig. 2 (c)). With the application of the electric field of +4 kV·cm$^{-1}$, magnetic domains cannot be switched at the same magnetic bias as the $V_o$-SRO (namely $H_{eff}$) has been erased as shown in Fig.

5 (c).

In summary, by designing an oxygen ionic chemical junction, $V_o$ are trapped at the interfacial-SRO phase, leading to an electronic and orbital reconstruction. As a result, a high-spin state at the $V_o$-SRO phase is achieved, which is reversibly switchable by an electric-field-induced oxygen migration in a solid-state ionic gating device. Furthermore, understanding of electronic and spin states in such a low-dimensional SRO phase makes this strongly correlated system a promising candidate for the further exploration and manipulation of emergent magnetism and spin structures.

## Acknowledgements

We gratefully acknowledge the discussions with Prof. Nicola A. Spaldin in Materials Theory of ETH Zürich and Dr. L. Wang in Center for Correlated Electron Systems of Institute for Basic Science. The work is supported by the National Key Research and Development Program of China through Contract 2016YFA0302300 and 2016YFA0300102 and 2017YFA0303602, the Basic Science Center Program of NSFC under Grant No. 51788104, the National Natural Science Foundation of China (Grant No. 11974052, 11675179, 11774360, 11904373), and Beijing Natural Science Foundation (Z190008) and the CAS Interdisciplinary Innovation Team. L. Si and K. Held were supported by the European Research Council (ERC) under the European Union's Seventh Framework Program (FP/2007-2013) through ERC Grant No. 306447, and by the Austrian Science Fund (FWF) through project P 30997. Beamtime from APS 33BM and SSRF 14B and 3315 Program of Ningbo are appreciated.

## Author contributions

J. Z. and Z. Z conceived the experiments and prepared the manuscript. X. Y. and C. T. fabricated the thin films. K.Z. and Y. S. prepared the SRO ceramic target. X. Y., Z. Lai, J. L., X. L. and C. T. performed the synchrotron x-ray diffraction, electrical transport and XPS measurements. X. Y. and J. W. carried out the SQUID measurement. L. S., P. J., Z. Z. and K. H. performed the theoretical calculations. I. M. and J. L. carried out the AFM and MFM measurements. Q. Z. and L. G. performed STEM measurement. Z. Luo performed the RSM measurement. J.Z., C.-W. N., Z. Z. and W. M. were involved in

**Main figures:**

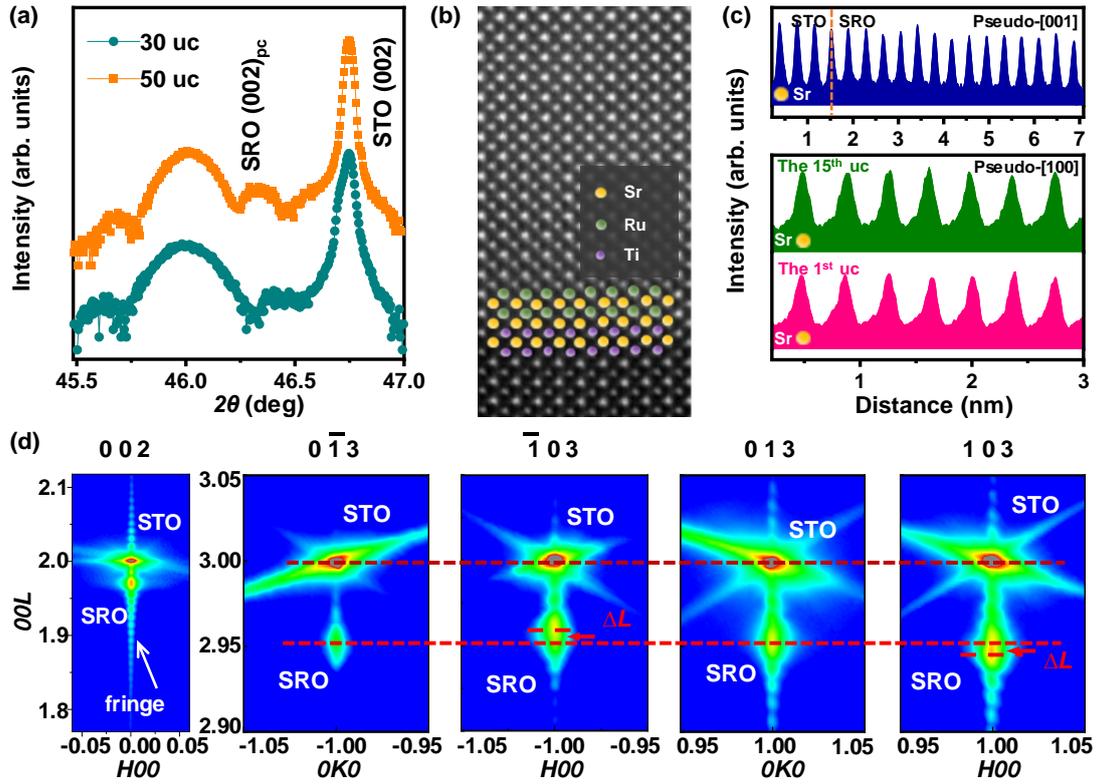

FIG. 1. High-quality Monoclinic SRO thin films without strain relaxation. (a) The $\theta$-$2\theta$ SXRD spectra for (002)-oriented sets of peaks in the 30 uc and 50 uc samples. The SRO thin film is the pseudocubic phase (pc). (b) High-resolution cross-section STEM image of 50 uc SRO thin film. (c) Line profiles of the SrO layers below 15 uc at the pseudo [001] orientation (top panel) and line profiles of the SrO layers for the 1st uc and the 15th uc of the SRO thin film (bottom panel). (d) RSM for 50 uc SRO on STO (001) around (002) and {103} STO Bragg reflection in the reciprocal lattice units of STO with the thickness fringes. The dissimilar values ($\Delta L$) indicate the monoclinic structure, where cell parameters measured for 50 uc SRO thin film are $a_{pc}$ = 3.905 Å, $b_{pc}$ = 3.905 Å, $c_{pc}$ = 3.967 Å, $\alpha$ = 90°, $\beta$ = 89.82° and $\gamma$ = 90°.

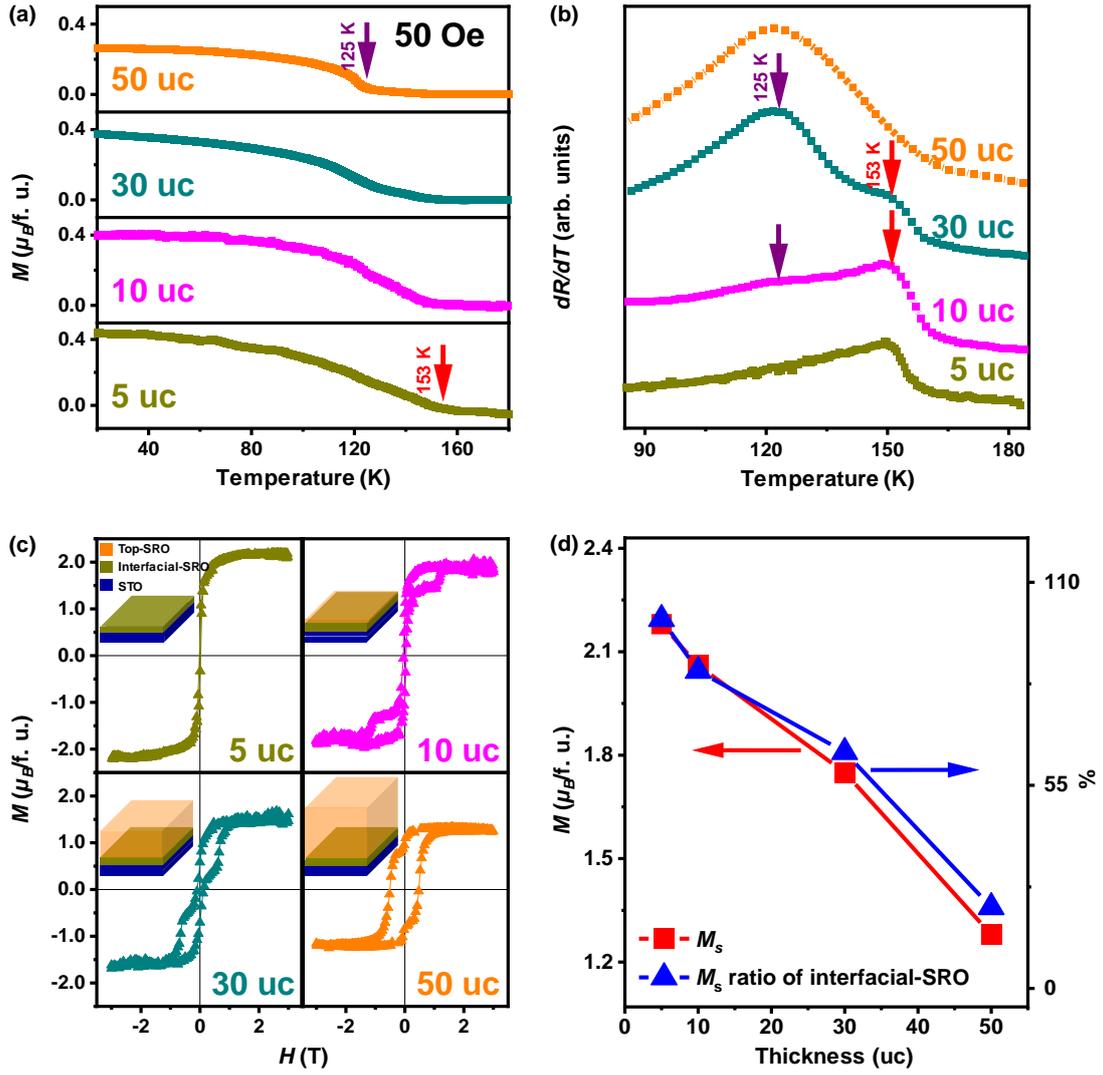

FIG. 2. Emergent interfacial magnetic phase with an enhanced magnetization. (a) Temperature-dependent magnetization (*M-T* curves) under a magnetic field of 50 Oe. The red and purple arrows indicate the $T_C$ of ~ 153 K and ~ 125 K, respectively. (b) Temperature-dependent differential resistivity (*dR/dT-T*) shows that the $T_C$ are consistent with the ones from *M-T* curves, indicating that the interfacial-SRO contributes to the high $T_C$. (c) Out-of-plane magnetic-field-dependent magnetization (*M-H* curves) at 10 K. Insets are the schematics of the interfacial-SRO and top-SRO phases. (d) Thickness-dependent out-of-plane $M_s$ (red curves) and its contribution from the interfacial-SRO (blue curves). As the thickness increases, the $M_s$ decreases, which indicates that the interfacial-SRO contributes to the enhancement of $M_s$. All the above data obtained from the monoclinic SRO thin films with thicknesses of 5 uc, 10 uc, 30 uc and 50 uc.

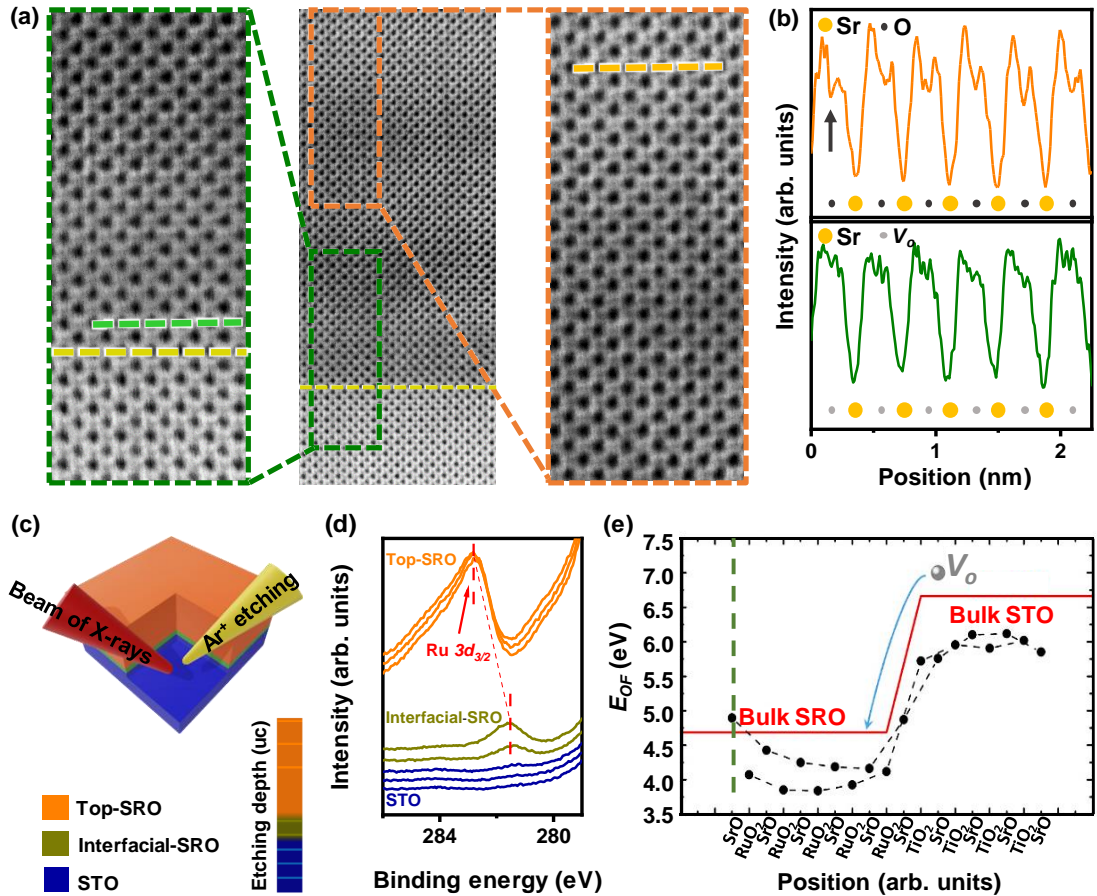

FIG. 3. Evidence of the interfacial-SRO phase driven by an ionic chemical junction. (a) Bright-field STEM of the SRO on STO. The left and right panels show the magnified bright-field STEM image as labeled by the green and orange dashed rectangle box in the middle panel. (b) The corresponding line profile for the detailed atomic structure of SrO layers in the interfacial-SRO (green curve) and top-SRO (orange curve). The yellow, dark grey and light grey dots represent the positions of Sr, O and $V_o$, respectively. The shallow valleys in the orange curve represent oxygen sites, which are negligible in the green curve, indicating more $V_o$ located at the interface. (c) Schematic of the depth-profile XPS measurements with *in-situ* Ar$^+$ etching. The orange, yellow and blue zones represent top-SRO, interfacial-SRO and STO, respectively. (d) Depth-profile binding energies of Ru $3d_{3/2}$ shows a dramatic change of binding energies for Ru $3d_{3/2}$. (e)

Calculation of the formation energies and diffusion barriers of $V_o$ across SRO/STO interface. The black dots are the formation energies $E_{OF}$ of the $V_o$ located at different layers across SRO/STO interface. The red lines are the bulk values of SRO and STO at $\sqrt{2} \times \sqrt{2} \times 4$ dimension. The grey sphere indicates the $V_o$ prefers to diffuse from STO into SRO and be prevented at the position of green dash line (diffusion length).

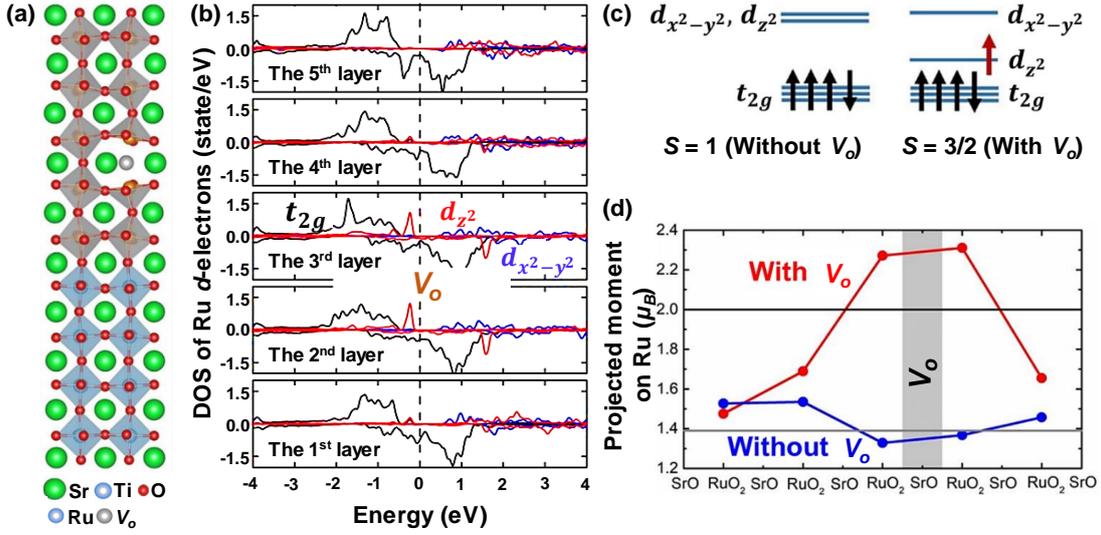

FIG. 4. Microscopic origin of the interfacial high-spin SRO. $V_o$-induced orbital reconstruction turns initial $4d^4$ low-spin state ($S = 1$) into $4d^5$ high-spin state ($S = 3/2$). In addition, the local magnetic moments of Ru sites neighboring the $V_o$ are significantly larger than that of other Ru sites far away from $V_o$. Moreover, they are beyond the upper limit of low-spin states predicted by Hund's rule for $Ru^{4+}$ ($t_{2g}^3 \uparrow$, $t_{2g}^1 \downarrow$, 2 $\mu_B$/Ru), which supports the emergence of high-spin state. (a) The crystal structure of SRO/STO heterostructure used in calculations, where four STO substrate layers and five SRO layers are constructed, and an oxygen atom is removed to model $V_o$. The green, light blue, orange, and red balls represent Sr, Ti, Ru, and O atoms, respectively. The $V_o$ is indicated by grey ball. (b) DFT-resulted layer-dependent DOS of Ru $d$-electrons. The $V_o$ is set in between the 2nd and 3rd $RuO_2$ layer. (c) The schematic illustrations of electronic configuration of low-spin state without $V_o$ and high-spin state with $V_o$. (d) Projected magnetic moments on Ru atoms with (red dots) or without (blue dots) $V_o$. The $V_o$ locates between the second and third $RuO_2$ layer as indicated by the grey area. The black and grey lines are the upper limit of low-spin state and projected moment on Ru atoms in bulk SRO, respectively.

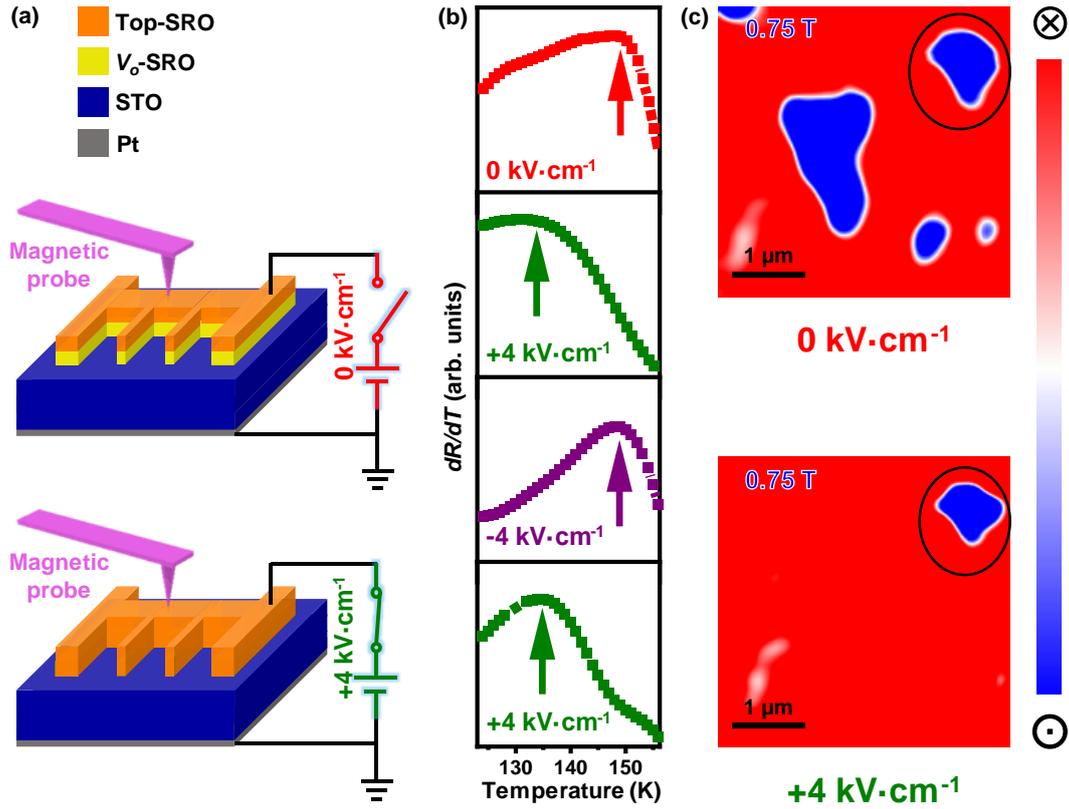

FIG. 5. Electric-field control of $T_C$ and mesoscopic domains switching. (a) Schematics of electrical transport under an electric field with *in-situ* MFM measurement in a solid-state ionic gating device. (b) Reversible control of $T_C$ derived from *dR/dT-T* curves (the red, green and purple arrows) due to the erasing and rebuilding of the $V_o$-SRO with the application of the vertical electric field of 0, +4, -4, +4 kV·cm$^{-1}$. (c) Magnetic domain switching in top-SRO/$V_o$-SRO (total thickness of 10 uc) without/with the application of electric field, where the magnetic domains were captured at 0.75 T and 10 K with/without $V_o$-SRO. The black circles indicate that the images were captured at the same area. The scale bar is 1 μm.